# Shape matters: DEM investigation of geometry-controlled mechanical response in irregular rock fragments under static and dynamic loading

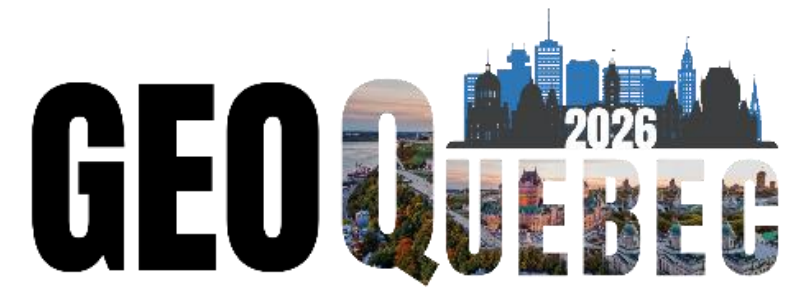

Athena Bahramiyarahmadi, Dr Rick Chalaturnyk
*University of Alberta, Edmonton, Alberta, Canada*

ABSTRACT
Mechanical characterization of subsurface rock formations typically requires standardized cylindrical core specimens, which are often unavailable from fractured or unconventional reservoir sequences. This study uses a Discrete Element Method (DEM) framework calibrated to Sulphur Mountain Formation siltstone to investigate how fragment geometry governs mechanical response in irregular rock particles, which are direct analogues of drill cuttings. Nineteen specimens (18 procedurally generated irregular geometries and one reference Brazilian disk), normalized to a 10 mm bounding sphere radius, were tested under quasi-static (Brazilian-type) and dynamic (Short Impact Load Cell) loading modes. Four mechanical outputs of failure force ($F_f$), apparent strength ($\sigma_f$) apparent stiffness ($K_a$), and stiffness (E) were regressed against seven retained shape descriptors. Results show that force-based quantities are strongly controlled by the surface-area-to-volume (SA/V) ratio under static loading (Adj. $R^2 \approx 0.69$), while area-normalized quantities ($\sigma_f$ and E) are geometrically insensitive. Dynamic loading amplifies geometric sensitivity for failure force and introduces an independent role for surface concavity depth. These findings establish a preliminary quantitative framework for interpreting mechanical measurements from irregular rock fragments when standardized core is unavailable.

RÉSUMÉ
La caractérisation mécanique des formations souterraines nécessite généralement des échantillons cylindriques normalisés, souvent indisponibles dans les réservoirs fracturés ou non conventionnels. Cette étude utilise un modèle basé sur la Méthode des Éléments Discrets (MED), calibré à partir de la siltite de la Formation du Mont Sulfure « Sulphur Mountain Formation » afin d'analyser l'influence de la géométrie des fragments sur la réponse mécanique des particules irrégulières de la roche, qui sont analogues directs des déblais générés lors du forage.
Dix-neuf échantillons ont été étudiés, dont 18 géométries irrégulières générées numériquement et un disque brésilien de référence, tous normalisés à un rayon de sphère de 10 mm. Les échantillons ont été soumis à des chargements quasi-statique (essai brésilien) ainsi qu'à des chargements dynamiques (essai par impact à courte durée). Les résultats obtenus pour quatre paramètres mécaniques à savoir la force de rupture, la résistance apparente, la rigidité apparente et rigidité, ont ensuite été corrélés à sept descripteurs géométriques de forme sélectionnés.
Les résultats montrent que les paramètres basés sur la force sont fortement contrôlés (influencés) par le rapport surface/volume (S/V) sous chargement statique, tandis que les paramètres normalisés par la surface (résistance et rigidité apparentes) demeurent globalement indépendants de la géométrie. Cependant, sous le chargement dynamique, la sensibilité géométrique de la force de rupture est davantage amplifiée, tout en mettant en évidence l'influence indépendante de la profondeur des concavités surfaciques. L'ensemble de ces observations permet de proposer un cadre quantitatif préliminaire pour interprétation des propriétés mécaniques des roches qui ont des formes irrégulières lorsque des carottes normalisées ne sont pas disponibles.

## 1 Introduction

Accurate geomechanical characterization of subsurface formations is a prerequisite for designing hydraulic fracture treatments, wellbore stability analyses, and reservoir stimulation programs in unconventional oil and gas plays (Xiao et al. 2024). The gold standard remains laboratory testing of cylindrical core specimens; however, core recovery from shale and siltstone reservoirs is frequently compromised by rock brittleness, in-situ stress unloading, and natural fracturing. Figure 1 presents an example of drill cores from unconventional resources in the western Canadian sedimentary basin, highlighting their highly-fractured state and the inapplicability of standard mechanical testing protocols to such material.

Drill cuttings represent a continuously available alternative sample stream. These small, irregularly shaped rock fragments are generated throughout the drilled interval and remain recoverable even in highly fractured formations. Their geometric nonconformity and relatively small sizes, however, renders classical testing protocols inapplicable: standard uniaxial, triaxial, and Brazilian tests presuppose idealized cylindrical geometries with controlled boundary conditions. Empirical approaches such as the Protodyakonov strength coefficient (Protodyakonov 1963), determined from a standardized drop-weight impact test on irregular rock fragments, and its subsequent refinements (e.g., Tsoi and Usol'tseva 2020; Ji et al. 2023), have long provided fragment-scale impact-based estimates of rock strength, but without treating fragment shape itself as a controlled or quantified variable. This study represents a preliminary attempt to investigate the potential of drilling cuttings for mechanical characterization of rocks through systematic DEM simulation.

The Discrete Element Method (DEM) provides a principled means to decouple geometric effects from intrinsic material behaviour (Cundall and Strack 1979; Potyondy and Cundall 2004). By holding bond-level micro-parameters constant across a geometrically diverse specimen library, shape becomes the sole experimental variable. This study leverages a DEM framework calibrated to Sulphur Mountain Formation siltstone (a stratigraphic equivalent of the Montney Formation) to evaluate whether mechanical outputs from irregular fragments can be predicted from their shape descriptors under both quasi-static and dynamic loading.

This paper: (i) characterizes 19 DEM specimens using seven shape descriptors; (ii) subjects all specimens to quasi-static (Brazilian-type) and dynamic (Short Impact Load Cell) loading; and (iii) evaluates statistical predictability of four mechanical outputs through simple linear regression, forward stepwise multiple regression, and AICc-based nonlinear model comparison.

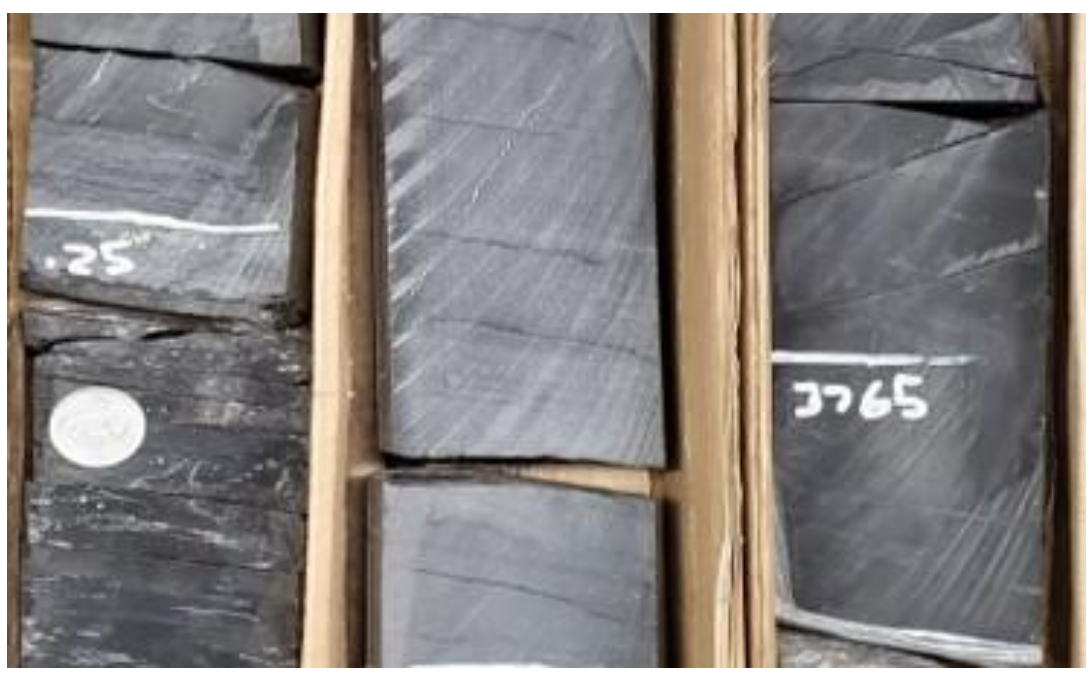

Figure 1. Example of highly-fractured drill cores from Duvernay Formation in Fox Creek area, AB., illustrating the compromised state of core material and the inapplicability of standard cylindrical specimen preparation for mechanical characterization.

## 2 Specimen geometry and shape descriptors

### 2.1 Geometry generation and normalization

Nineteen specimen geometries were produced using Blender 3.6's built-in Rock Generator add-on (Add Mesh: Extra Objects). Eighteen irregular fragments were created by parametric mesh distortion of spherical seed surfaces, followed by random vertex perturbation and smoothing to yield geometrically plausible, manifold-closed volumes. One reference specimen - a Brazilian disk (rock_019, diameter 20 mm, height 10 mm) - serves as the benchmark. To isolate shape effects from scale effects, all irregular specimens were normalized to a uniform bounding sphere of 10 mm radius using the Miniball algorithm (Welzl 1991).

Prior to testing, each specimen underwent a quasi-static gravitational settlement to identify its most stable resting orientation.

### 2.2 Shape descriptor quantification

Twenty-one shape descriptors were initially computed per specimen. An inter-correlation screen (Pearson |r| > 0.85) reduced this to seven non-redundant descriptors: (1) SA/V ratio ($mm^{-1}$); (2) Zingg c/a ratio; (3) moment ratio $I_3/I_1$; (4) RMS surface roughness Sq (mm); (5) Wadell roundness R (Wadell 1932); (6) depth of concavity (mm); and (7) Masad angularity index AI (Masad et al. 2001).

Figure 2 shows the Pearson inter-correlation matrix for the seven retained descriptors (n = 19 specimens). All pairs satisfy $|r| \leq 0.82$, confirming descriptor independence and ruling out multicollinearity in the regression models. The strongest inter-descriptor correlation is between SA/V ratio and depth of concavity ($r = -0.77$), reflecting the physical link between surface compactness and re-entrant geometry. Table 1 summarizes statistics across all 19 specimens. The SA/V ratio spans 0.36–0.70 $mm^{-1}$, reflecting morphological diversity.

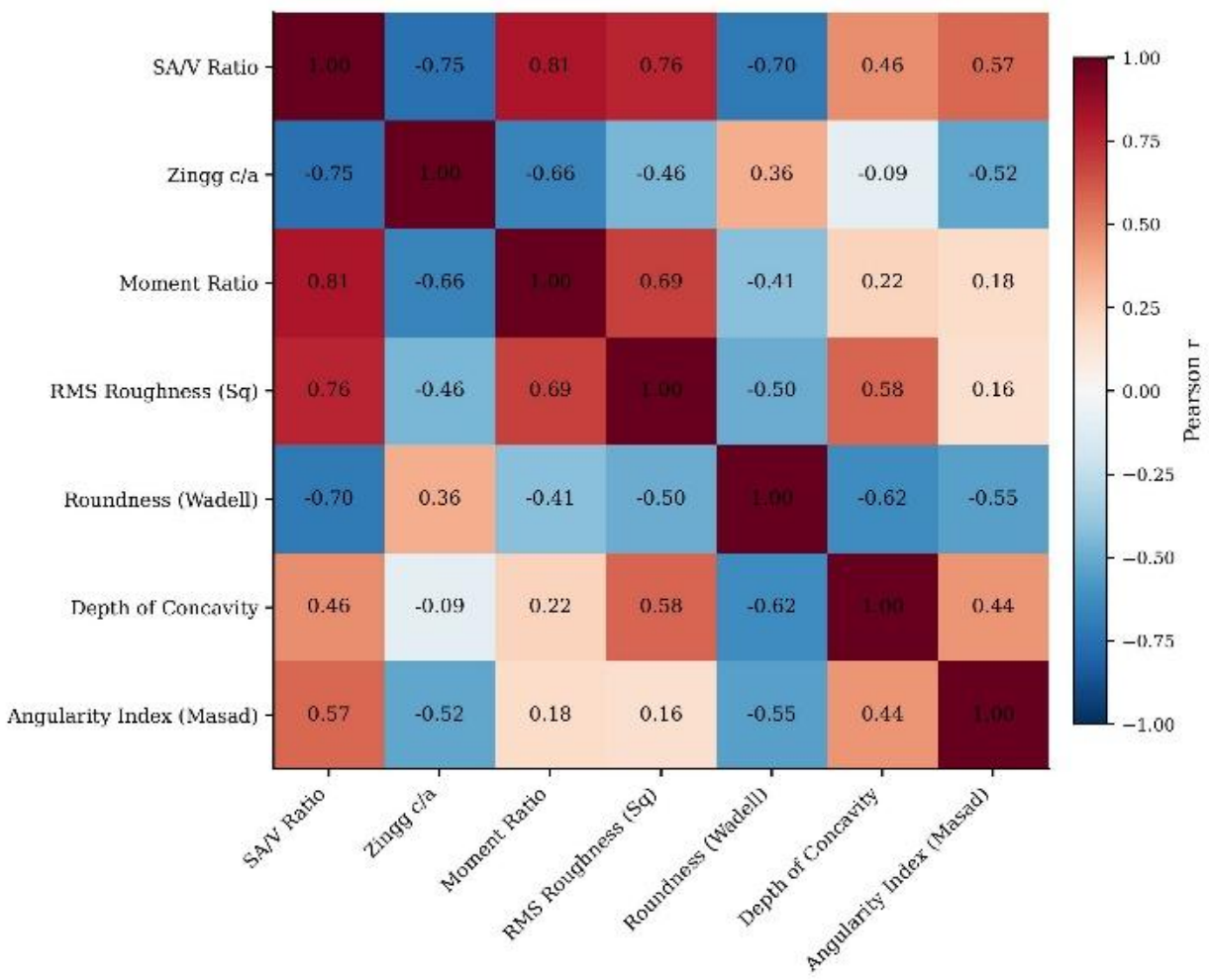

Figure 2. Pearson inter-correlation matrix for the seven retained shape descriptors (n = 19). Cell values show r. All pairs satisfy $|r| \leq 0.82$, confirming descriptor independence.

## 3 Numerical methodology

### 3.1 DEM framework and model calibration

All simulations were conducted using ESyS-Particle, a high-performance open-source DEM framework (Wang et al. 2006). The Bonded Particle Model (BPM) represents rock as a dense packing of spheres connected by elastic-brittle beam bonds capable of transmitting forces and moments in six degrees of freedom (Potyondy and Cundall 2004). Bond failure follows a Mohr–Coulomb criterion: a bond severs when shear stress exceeds cohesion plus normal stress times friction angle. Bond micro-parameters were calibrated against UCS and Brazilian tensile strength data from 10 mm siltstone specimens of the Sulphur Mountain Formation (Banff area, Alberta) using a systematic surrogate modelling approach.

### 3.2 Loading configurations and output derivation

Two loading regimes were applied to all 19 specimens: (i) Static loading: diametrical compression at 0.001 mm/ms, replicating the Brazilian tensile test; and (ii) Dynamic loading: impact replicating the Short Impact Load Cell (SILC), in which a steel ball free-falls onto the specimen resting on a long steel rod, targeting high-strain-rate responses relevant to comminution (King and Bourgeois 1993; Tavares and King 1998).

Four mechanical outputs were extracted, failure force $F_f$ (kN): peak platen contact force; apparent strength $\sigma_f$ (MPa): $F_f$ normalized by the square of the initial loading-point separation ($h_0$) (Jaeger 1967); apparent stiffness $K_a$ (kN): force - axial strain slope. The axial strain was defined as $\varepsilon=\Delta u/h_0$, where $\Delta u$ is the relative wall displacement; and stiffness E (GPa): slope of $\sigma_f$ - axial strain.

Figure 3 and Figure 4 show representative force – axial strain histories for selected specimens under both loading modes, illustrating how geometric irregularity modulates the loading path, peak force, and post-failure unloading behaviour.

### 3.3 Failure pattern analysis

To qualitatively assess how geometry and loading rate govern fracture mechanics, the evolution of mechanical damage was tracked for all specimens. The evolution of damage for three representative specimens: Rock_004 (elongated, angular; SA/V = 0.470 mm⁻¹), Rock_011 (smooth and convex; SA/V = 0.362 mm⁻¹, Rnd = 1.000, AI = 0.018, the least angular irregular specimen in the library), and the reference cylinder Rock_019 (SA/V = 0.401 mm⁻¹) under static and dynamic loading conditions are presented in Figure 5 and Figure 6. Bonds are coloured by strain magnitude: cool tones (blue) indicate compression zones; warm tones (red/orange) indicate tensile zones.

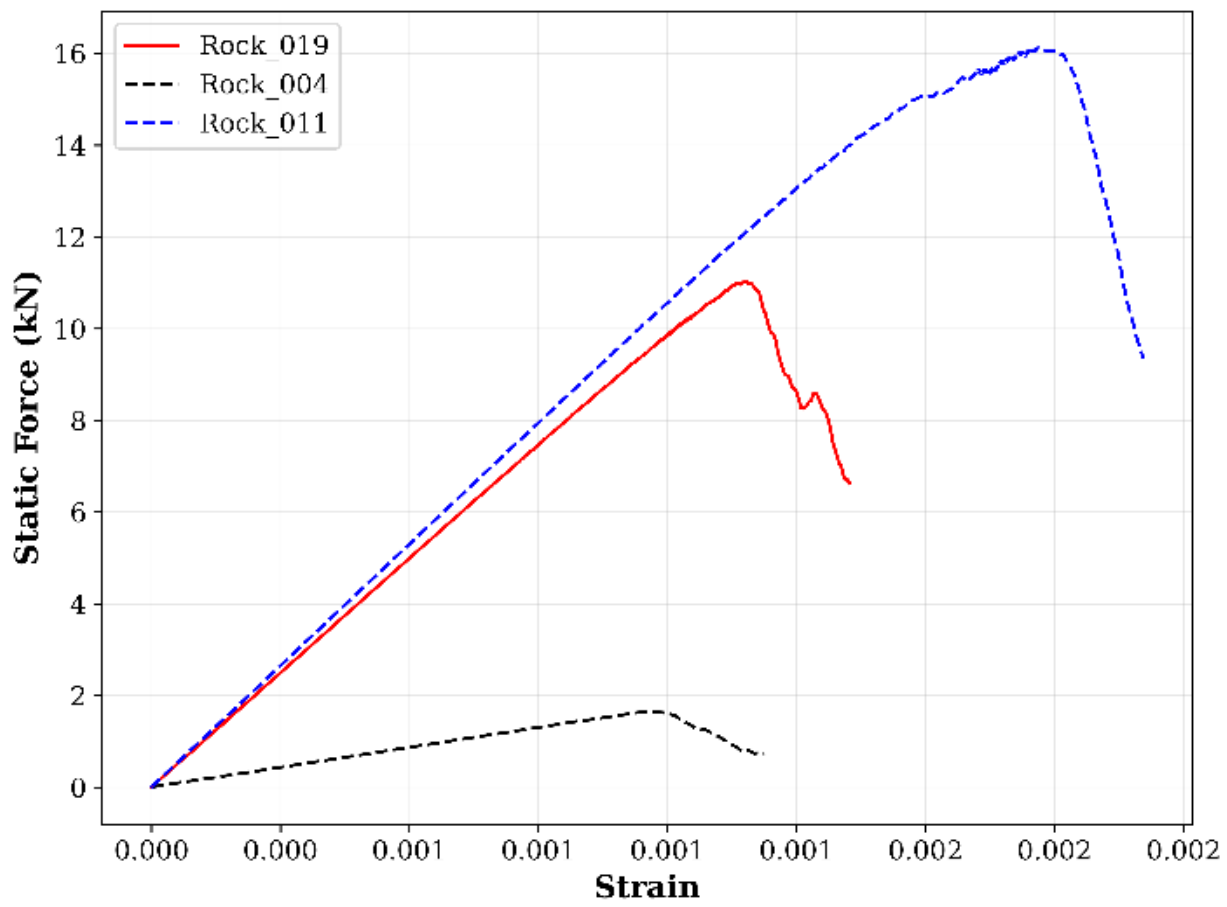

Figure 3. Force – strain histories for selected specimens under static loading. The reference cylinder rock_019 (red) provides a standardized benchmark.

Under static loading (Figure 5), all three specimens develop compressive fractures initiating at contact points. Rock_019 displays a clean, near-diametral tensile split consistent with classical Brazilian test behaviour. Rock_004's elongated geometry creates preferential tensile zones along the loading axis, producing a planar but asymmetric fracture. Rock_011, with its near-spherical convex surface, exhibits more stable compressive zone within the specimen's core, which shifts the critical tensile strains toward the periphery, causing the sample to fail from the outside in, where the tensile strain is maximized.

The failure patterns observed in these specimens under dynamic loading are broadly consistent with those under static loading. The main difference is the strain magnitude associated with failure. Although the strain concentration pattern in the dynamic regime is similar to that observed in the static regime, failure occurs at a substantially higher strain magnitude under dynamic loading.

Table 1. Summary statistics of the seven retained shape descriptors (n = 19 specimens).

| Descriptor | Unit | Mean ± SD | Range | rock-019 |
|---|---|---|---|---|
| SA/V ratio | mm⁻¹ | 0.46 ± 0.09 | 0.36–0.70 | 1 |
| Zingg c/a ratio | – | 0.55 ± 0.10 | 0.41–0.75 | 60 |
| Moment ratio | – | 1.89 ± 0.51 | 1.22–3.54 | 44 |
| RMS roughness | mm | 1.32 ± 0.41 | 0.00–1.91 | 18 |
| Roundness R | – | 0.97 ± 0.03 | 0.87–1.00 | 8 |
| Depth of concavity | mm | 3.43 ± 2.18 | 0.00–5.83 | 44 |
| Angularity index AI | – | 0.09 ± 0.05 | 0.02–0.19 | 0.17[1] |

[1] AI of rock_019 reflects a mesh discretization artefact: the cylinder STL is composed of a finite number of planar polygonal faces, yielding an elevated value of 0.17 despite representing an idealised smooth surface.

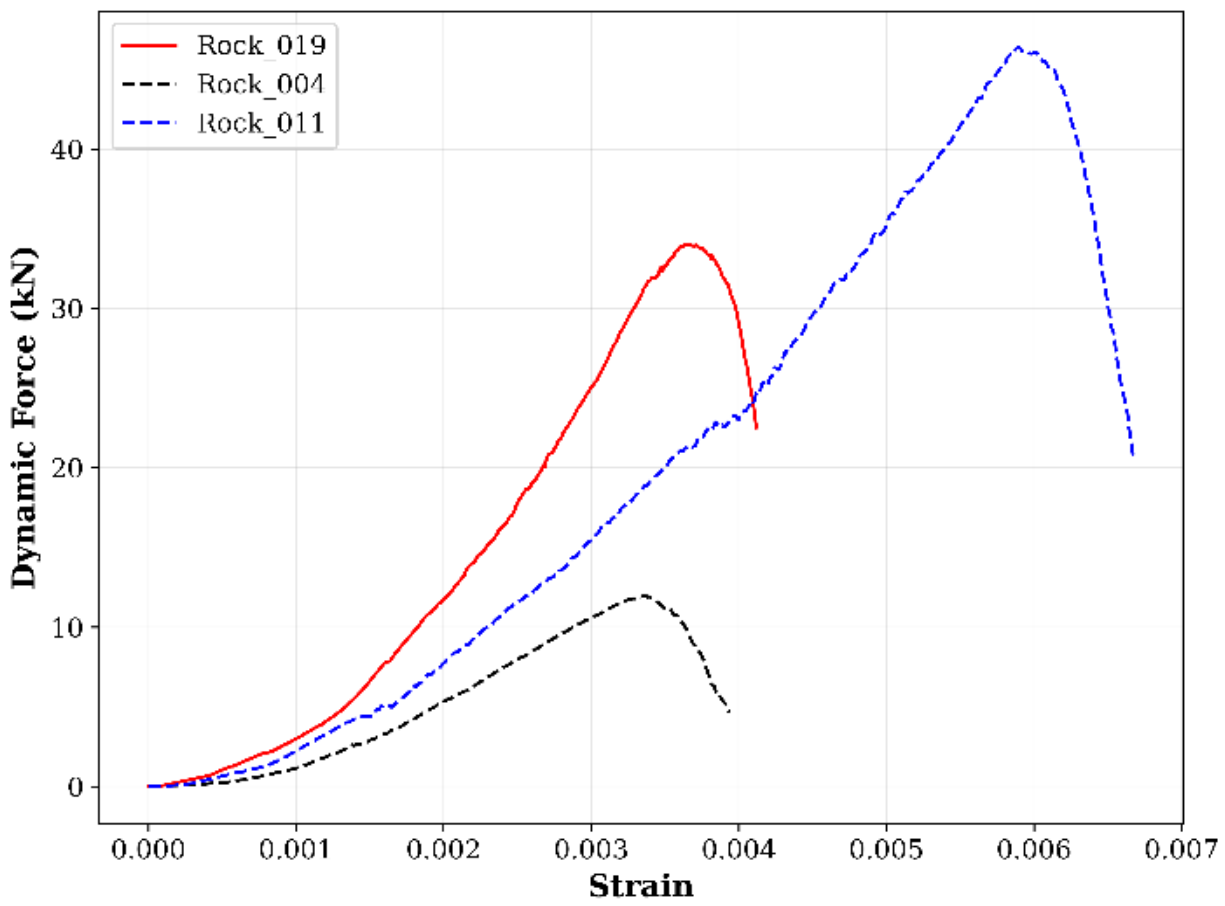

Figure 4. Force – strain histories for selected specimens under dynamic loading. The reference cylinder rock_019 (red) provides a standardized benchmark.

# 4 Results and discussion

## 4.1 Mechanical response overview

All four mechanical outputs exhibit substantial inter-specimen variability under both loading modes. $F_{f\text{-static}}$ ranges from 1.60 to 16.13 kN; $F_{f\text{-dynamic}}$ ranges from 6.05 to 46.36 kN, a roughly three-fold amplification consistent with rate-strengthening in brittle rock (Mahanta et al. 2018). The reference cylinder (rock_019) consistently occupies the mid-to-upper force range, reflecting its compact, symmetric geometry. The force–axial strain histories (Figure 3 and Figure 4) further illustrate how peak force, pre peak stiffness, and post-failure unloading differ systematically between compact and irregular specimens under each loading regime.

Area-normalized quantities ($\sigma_{f\text{-static}}$ and $E_{static}$) in static mode display considerably reduced inter-specimen scatter relative to force-based outputs. This reduction in scatter has physical meaning and also occurs under dynamic loading, though in the dynamic case, the scatter levels are roughly comparable between the different normalized parameters, with one showing slightly greater variability. Nonetheless, across both loading modes, one particular normalized metric exhibits the smallest degree of scatter.

Normalizing the measured force by a geometry-dependent area effectively removes much of the shape-related influence, resulting in only a minor residual geometric effect in the normalized parameters.

## 4.2 Shape – property correlations

To explore possible relationships between geometric shape characteristics and the measured mechanical properties, several regression approaches were employed, including simple linear regression, forward stepwise multiple linear regression (MLR), and evaluation of nonlinear regression models.
Simple linear regressions were computed for all descriptor–output pairs.

### 4.2.1 Simple linear regression

Figure 7 presents $R^2$ values for static and dynamic loading side by side across all selected descriptor–output pairs. Three findings emerge: (1) the qualitative descriptor hierarchy is preserved across loading modes (SA/V ratio dominates $F_f$ in both loading modes; $\sigma_f$ and E remain poor throughout), confirming that the link between geometric compactness and load-bearing capacity is robust to loading rate; (2) dynamic geometric sensitivity is amplified: the SA/V - $F_f$ regression slope is approximately three times steeper under dynamic loading, consistent with stress-wave amplification sensitivity to strain rate at geometric irregularities; and (3) the dominant $K_a$ predictor shifts from SA/V ratio (static, $R^2$ = 0.67) to RMS roughness (dynamic, $R^2$ = 0.60) indicating that surface micro-topography controls stiffness under inertial loading when quasi-static force equilibration no longer applies.

### 4.2.2 Forward stepwise MLR

Since the seven retained shape descriptors are not fully orthogonal to one another (Figure 2), fitting all possible subsets simultaneously would introduce multicollinearity and make individual coefficient estimates unreliable. Forward stepwise selection was therefore used to build parsimonious models one predictor at a time

Table 2 summarizes the final MLR models, selected by forward stepwise selection for each mechanical output and loading mode.

Under static loading, the stepwise procedure confirmed that $F_f$ and $K_a$ are each well-described by a single descriptor. For failure force, the SA/V ratio alone accounts for 69.4% of the variance after the degrees-of-freedom correction (Adj. $R^2$ =0.69), and no second descriptor could improve upon this once the variance it shared with SA/V ratio had been removed. The final model predicts the failure force of an unseen specimen with an expected error of roughly 60% of the population spread (3.95 kN). For the area-normalized outputs, the static MLR results paint a more cautionary picture. Both $\sigma_f$ and E accept a two-predictor combination of moment ratio and RMS roughness, but the resulting Adj. $R^2$ values of 0.252 and 0.236 represent only a modest step above the noise floor. Under dynamic loading, the most significant departure from the static results is the acceptance of a two-predictor model for $F_f$. After SA/V ratio is entered at Step 1 (Adj. $R^2$ =0.60), depth of concavity is admitted at Step 2 with a partial p=0.007 and a substantial improvement in Adj. $R^2$ of +0.14, bringing the final model to Adj. $R^2$ =0.74. For the remaining dynamic outputs, the stepwise procedure terminates at Step 1 in all three cases.

### 4.2.3 Nonlinearity assessment

AICc model selection (Burnham and Anderson 2002) is used to compare linear, power law, and quadratic polynomial fits for each best-predictor–output pair.

Under static loading, the linear model is selected for $F_f$ and $K_a$ ($\Delta$AICc < 2 for alternatives). For $\sigma_f$ and E, the polynomial is selected ($\Delta$AICc > 2), though absolute $R^2$ remains below 0.38.

Under dynamic loading, the power law supersedes the linear model for $F_f$ vs SA/V ($\Delta$AICc = 6.27) in favour of power law), indicating nonlinear geometric sensitivity: as irregularity increases, $F_{f\text{-dynamic}}$ declines at an accelerating rate consistent with stress-wave amplification that scales nonlinearly with surface compactness. For $K_a$, $\sigma_f$, and E under dynamic conditions, the linear model is adequate. Figure 8 presents the model comparison for $F_f$ under both static (top panels) and dynamic (bottom panels) loading.

# 5 Summary and conclusions

This study represents a preliminary attempt to investigate the potential of drilling cuttings for mechanical characterization of rocks, demonstrating that fragment shape systematically governs force-based mechanical outputs under both quasi-static and dynamic loading. The principal conclusions are:

1. Force-based outputs ($F_f$ and $K_a$) are predictable from shape descriptors under both loading modes ($R^2$ values up to 0.71 for $F_{f\text{-static}}$ and 0.76 for $F_{f\text{-dynamic}}$) with the SA/V ratio as the dominant predictor.

2. Area-normalized quantities ($\sigma_f$ and E) are effectively geometrically insensitive, meaning the effective area normalization cancels the shape signal, and cannot be reliably predicted from shape descriptors at n = 19.
3. Dynamic loading amplifies geometric sensitivity in $F_f$ by approximately three-fold and introduces an independent predictive role for surface concavity depth ($\Delta$Adj. $R^2$ = +0.14) at Step 2 of the $F_{f\text{-dynamic}}$ model).
4. Failure patterns under dynamic loading are similar to the static mode, yet in dynamic loading higher strain magnitudes result in failure of the specimen.
5. Linear models are adequate for description of $F_{f\text{-static}}$ and $K_{a\text{-static}}$; a power law better describes the $F_{f\text{-dynamic}}$ - SAV relationship ($\Delta$AICc =6.27 vs linear); reflecting nonlinear stress - geometry interaction.

These findings lay the groundwork for a shape-informed interpretation framework for drill cuttings-based mechanical characterization. Future work expands the geometric library, validates against physical impact data, and investigates potential methods to eliminate shape-induced variability in mechanical characterization of rocks.

## 6 Acknowledgements

The authors gratefully acknowledge the computational resources provided by the Digital Research Alliance of Canada and the support of the Reservoir Geomechanics Research Group staff at the University of Alberta. The authors also extend special thanks to Dr. Dion Weatherley for his invaluable guidance and support in the use of the ESyS-Particle DEM framework.

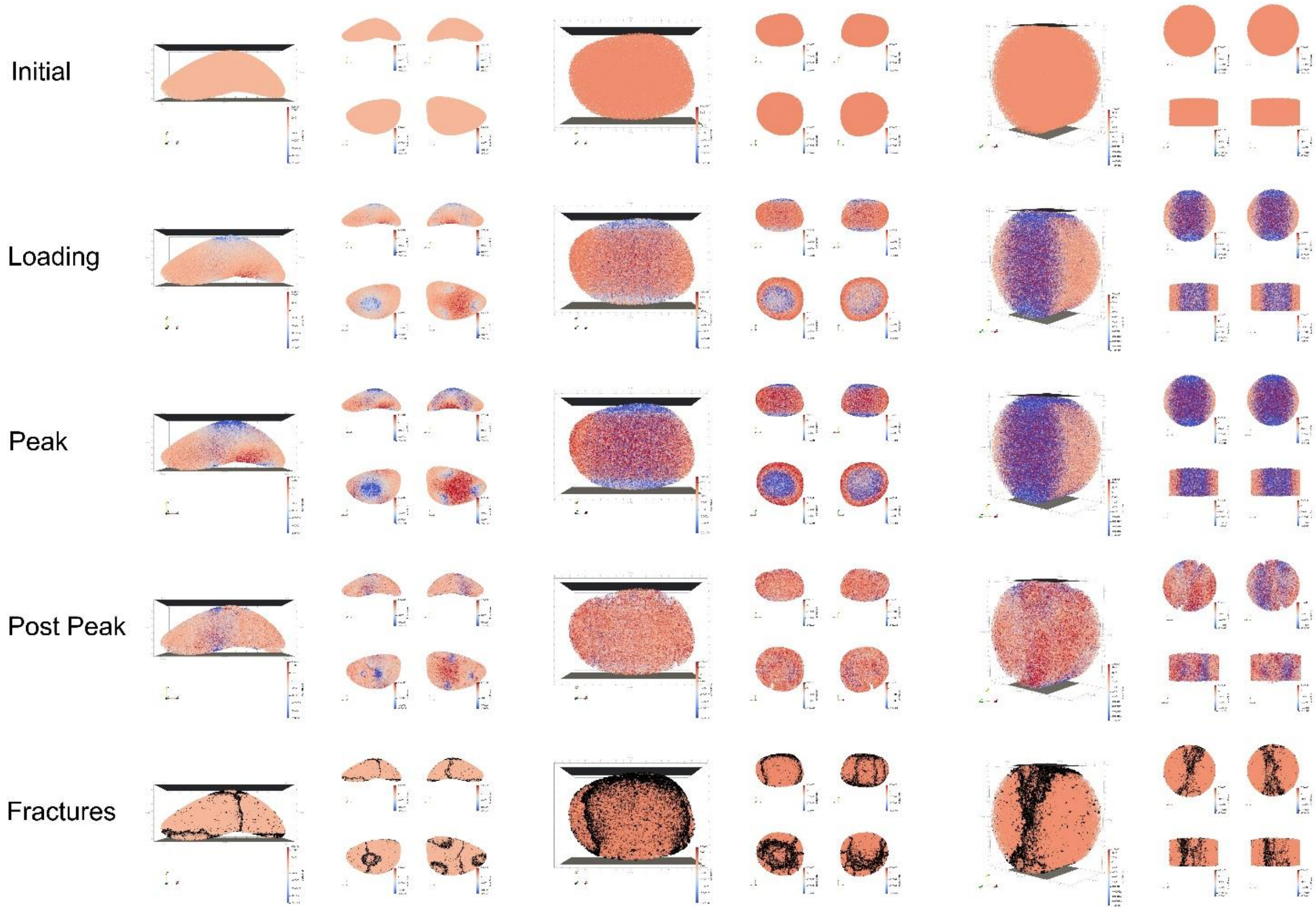

Figure 5. Evolution of mechanical damage and bond strain distribution under static loading for Rock_004 (left), Rock_011 (centre), and Rock_019 reference cylinder (right). Cool colours (blue) denote compression; warm colours (red/orange) denote tensile zones. Rock_019 exhibits a clean diametral split; Rock_011 shows concentrated central splitting owing to its smooth convex surface; Rock_004 develops an asymmetric planar fracture network.

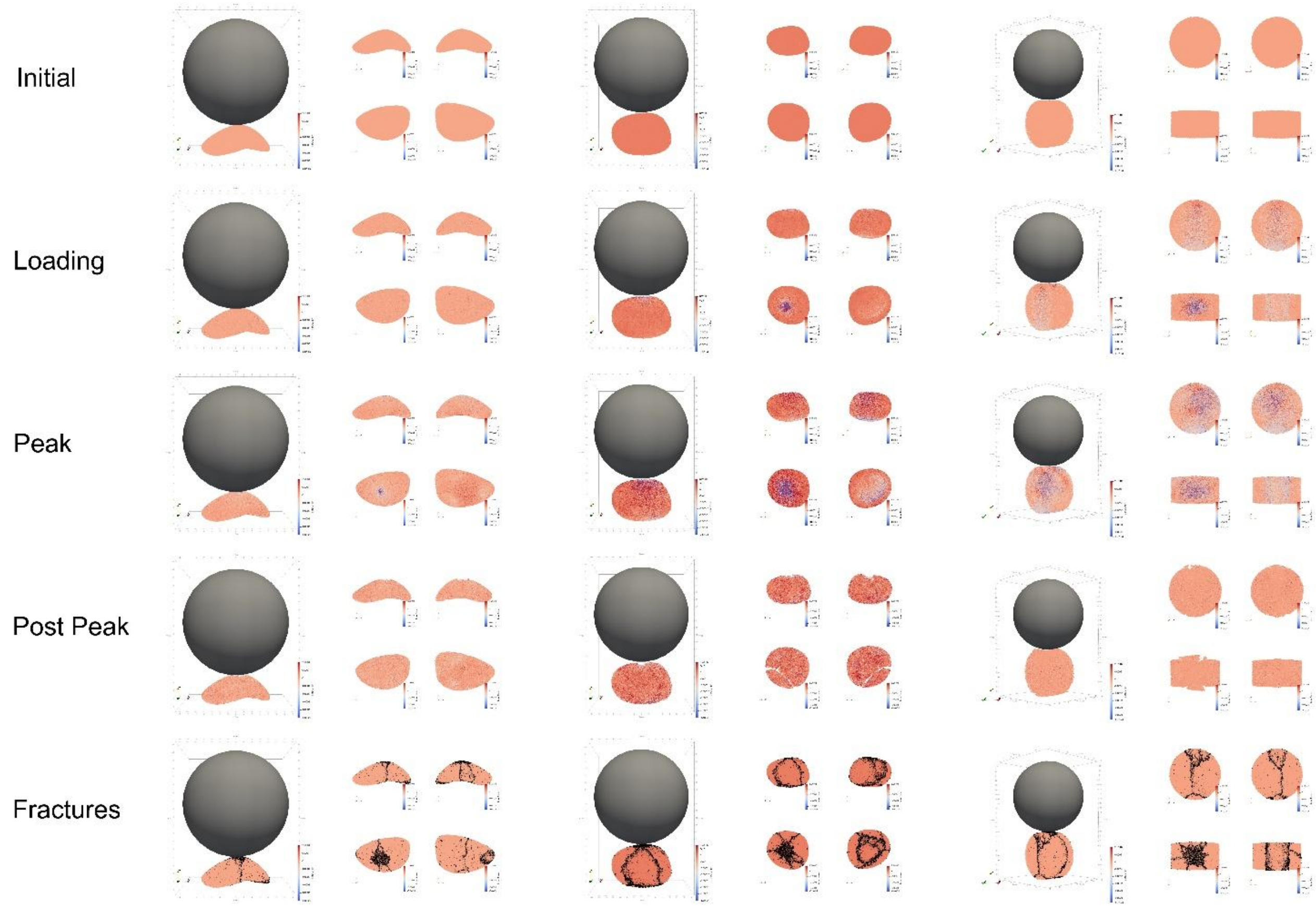

Figure 6. Evolution of mechanical damage and the distribution of bond strain during dynamic loading conditions for the same three specimens. The failure modes observed under both loading scenarios show comparable characteristics

Table 2. Final multiple linear regression models selected by forward stepwise selection for each mechanical output and loading mode. $R^2$, Adj. $R^2$, and LOO-CV RMSE are reported for the final model.

| Parameter | Mode | Step 1 Predictor | Step 2 Predictor | $R^2$ | Adj. $R^2$ | LOO-CV RMSE | Std.Dev. |
|---|---|---|---|---|---|---|---|
| $F_f$ (kN) | static | SA/V Ratio | - | 0.71 | 0.69 | 2.33 | 3.95 |
| | dynamic | SA/V Ratio | Depth of Concavity | 0.76 | 0.74 | 5.58 | 9.54 |
| $\sigma_f$ (MPa) | static | Moment Ratio | RMS Roughness | 0.33 | 0.25 | 27.90 | 28.58 |
| | dynamic | Depth of Concavity | - | 0.07 | 0.02 | 73.34 | 66.66 |
| $K_a$ (kN) | static | SA/V Ratio | - | 0.67 | 0.65 | 1701.88 | 2699.32 |
| | dynamic | RMS Roughness | - | 0.60 | 0.58 | 2083.28 | 3055.72 |
| E (GPa) | static | Moment Ratio | RMS Roughness | 0.32 | 0.24 | 18.64 | 18.73 |
| | dynamic | RMS Roughness | - | 0.19 | 0.15 | 24.83 | 24.93 |

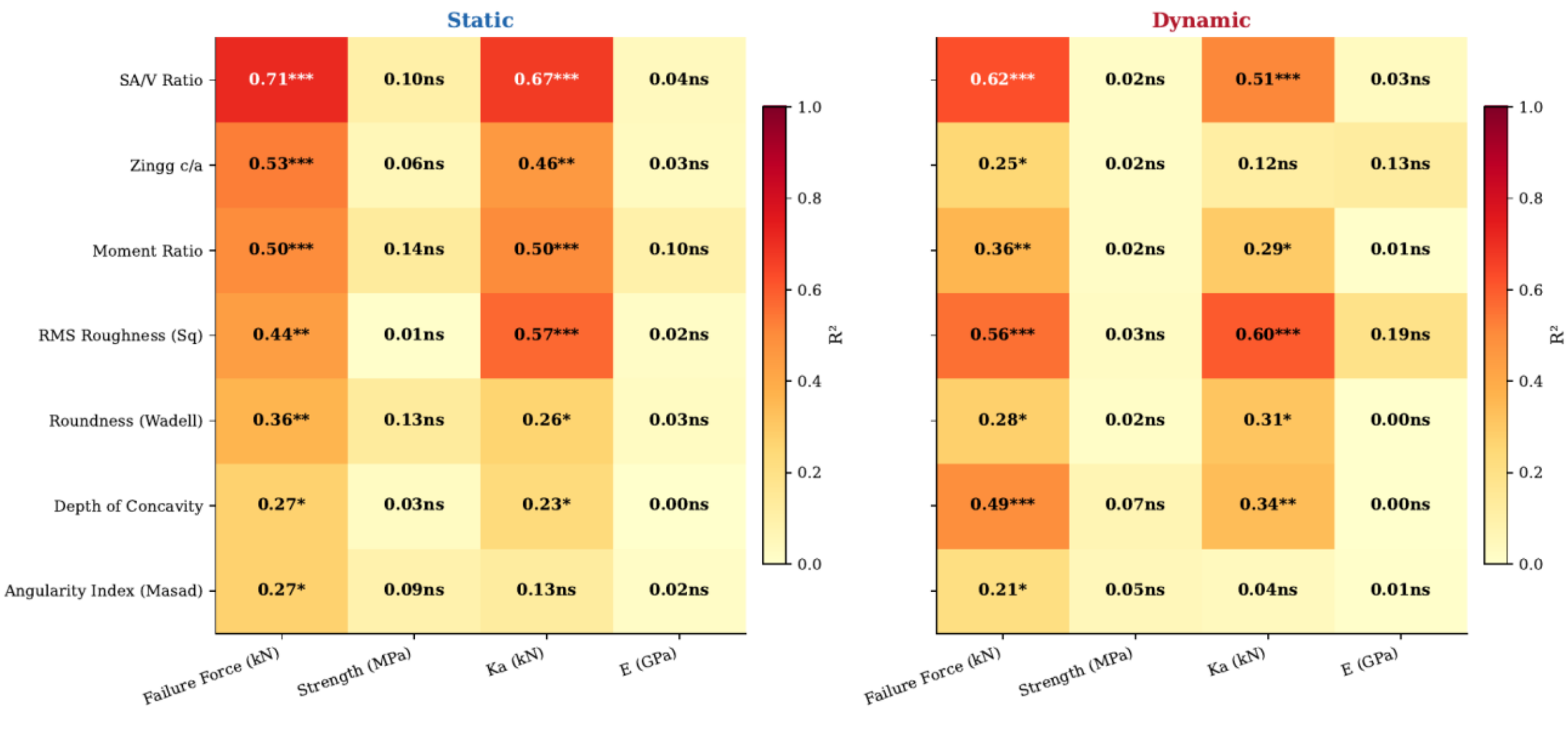

Figure 7. $R^2$ comparison between static (left panel) and dynamic (right panel) simple linear regressions. Cell text shows $R^2$ with significance stars (* $p < 0.05$, ** $p < 0.01$, *** $p < 0.001$). Shape sensitivity is preserved for force-based outputs across loading modes; area-normalized outputs remain insensitive in both regimes.

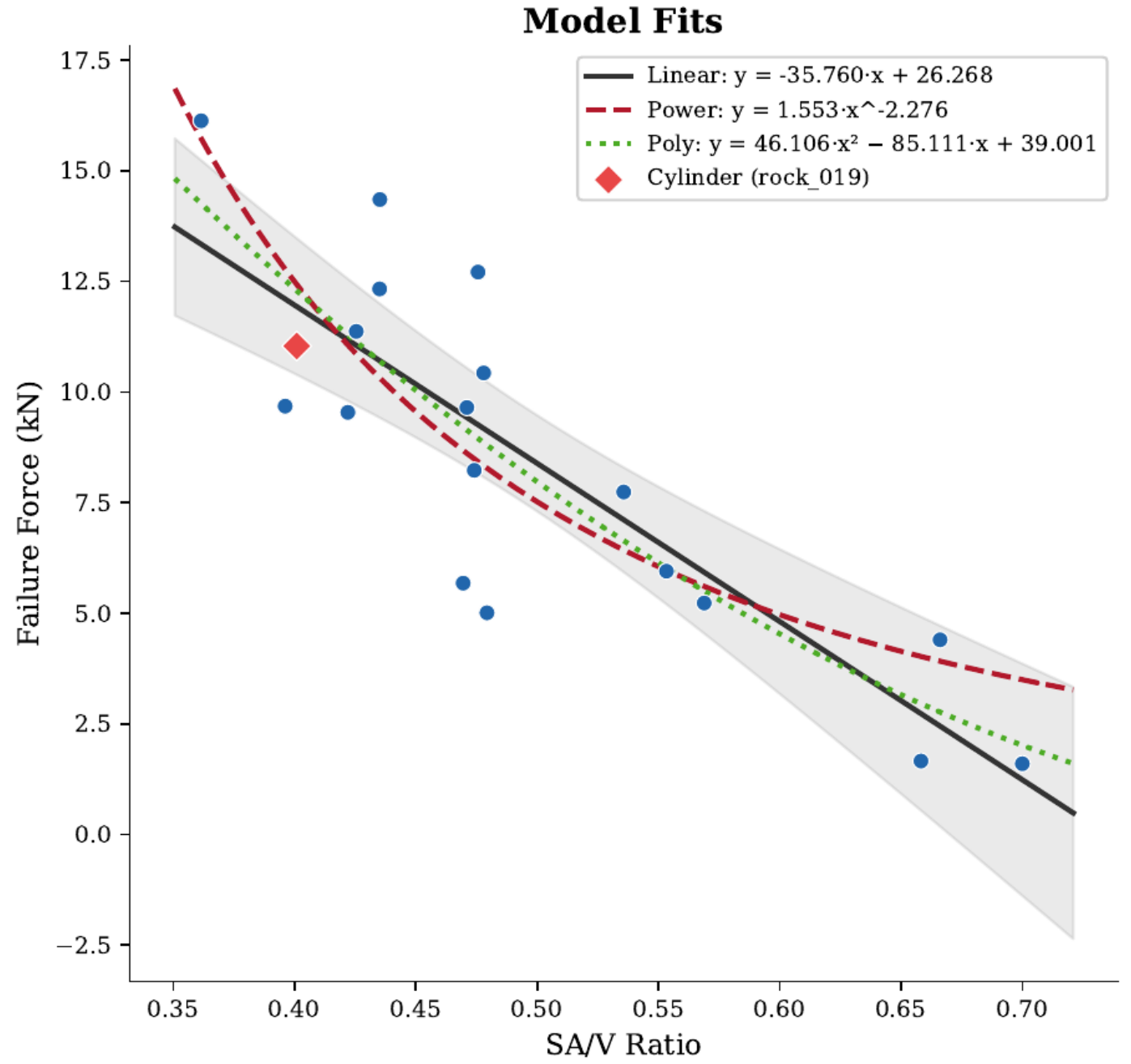

**Model Comparison (green = best AICc)**

| Model | k | $R^2$ | Adj.$R^2$ | AICc | ΔAICc |
|---|---|---|---|---|---|
| **Linear** | **2** | **0.711** | **0.694** | **33.32** | **0.00** |
| Power Law | 2 | 0.698 | 0.680 | 34.18 | 0.86 |
| Polynomial | 3 | 0.722 | 0.687 | 35.47 | 2.15 |

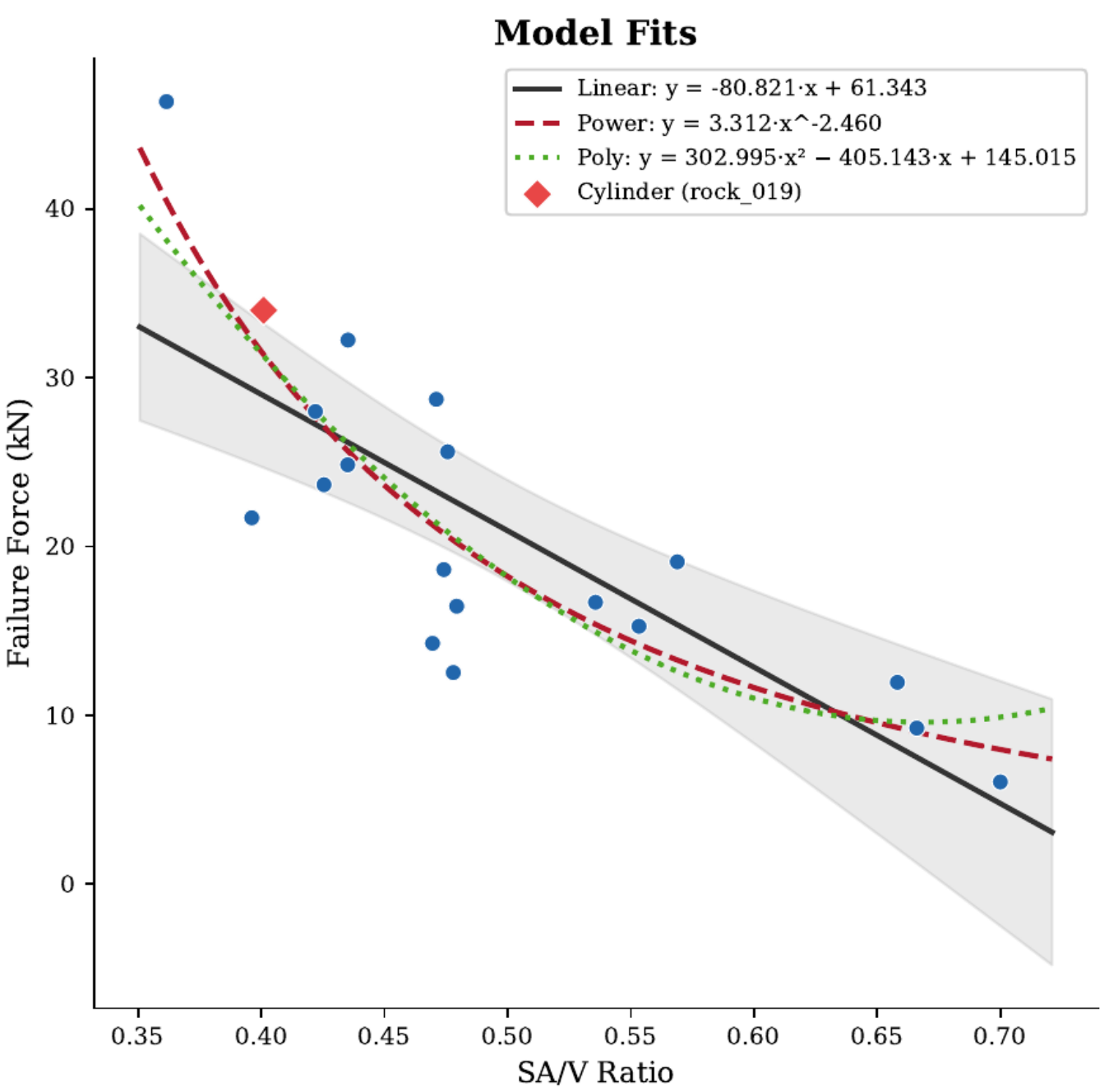

**Model Comparison (green = best AICc)**

| Model | k | $R^2$ | Adj.$R^2$ | AICc | ΔAICc |
|---|---|---|---|---|---|
| Linear | 2 | 0.622 | 0.600 | 71.99 | 6.27 |
| **Power Law** | **2** | **0.728** | **0.712** | **65.72** | **0.00** |
| Polynomial | 3 | 0.699 | 0.661 | 70.51 | 4.79 |

Figure 8. Nonlinear model comparison for failure force ($F_f$) vs SA/V ratio under static loading (top row) and dynamic loading (bottom row). Left panel: scatter with three fitted curves (grey solid = linear, red dashed = power law, green dotted = polynomial) and shaded 95% CI for the linear fit. Right panel: AICc comparison table; green row = selected model. Under static conditions the linear model wins; under dynamic conditions the power law is clearly superior (ΔAICc =6.27).